\documentclass[conference]{IEEEtran}
\IEEEoverridecommandlockouts

\usepackage{cite}
\usepackage{amsmath,amssymb,amsfonts}
\usepackage{algorithmic}
\usepackage{graphicx}
\usepackage{textcomp}
\usepackage{xcolor}
\usepackage{amsmath, amssymb, amsfonts}
\usepackage{algorithmic}
\usepackage{graphicx}
\usepackage{textcomp}
\usepackage{xcolor}
\usepackage{cite}
\usepackage[most]{tcolorbox}
\usepackage{url}
\usepackage{mdframed}
\usepackage{tcolorbox}
\usepackage{enumitem}
\def\BibTeX{{\rm B\kern-.05em{\sc i\kern-.025em b}\kern-.08em
    T\kern-.1667em\lower.7ex\hbox{E}\kern-.125emX}}
\usepackage{tabularx, array, booktabs, multirow, multicol, float, stfloats}
\usepackage[flushleft]{threeparttable}
\usepackage{adjustbox}
\usepackage{hyperref}
\usepackage{pgf, tikz, pgfplots}
\usetikzlibrary{patterns, arrows.meta, calc, positioning, shapes, decorations.markings}
\pgfplotsset{compat=newest}

\usepackage{caption}
\usepackage{subcaption}

\usepackage{soul, csquotes}
\usepackage[table]{xcolor}
\definecolor{codegreen}{rgb}{0,0.6,0}
\definecolor{codegray}{rgb}{0.5,0.5,0.5}
\definecolor{codepurple}{rgb}{0.58,0,0.82}
\definecolor{backcolour}{rgb}{1,1,1}
\definecolor{MyBlue}{rgb}{0.0,0.5,1.0}
\definecolor{MyRed}{rgb}{1.0,0.0,0.0}
\usepackage[table]{xcolor}
\definecolor{softgray}{gray}{0.95}
\usepackage{listings}
\newtcolorbox{promptbox}{
  colback=white,
  colframe=white!0,
  boxrule=0.5pt,
  arc=4pt,
  left=10pt,
  right=10pt,
  top=10pt,
  bottom=10pt,
  fontupper=\small,
  before skip=10pt,
  after skip=10pt,
  breakable
}

\newtcolorbox{promptboxx}{
  colback=white,
  colframe=black!50,
  boxrule=0.5pt,
  arc=4pt,
  left=10pt,
  right=10pt,
  top=10pt,
  bottom=10pt,
  fontupper=\small,
  before skip=10pt,
  after skip=10pt
}

\newcommand{\codeblock}[1]{
  \begin{tcolorbox}[colback=gray!5, colframe=blue!50!black, boxrule=0.5pt, arc=4pt, fontupper=\ttfamily\small, left=5pt, right=5pt]
    #1
  \end{tcolorbox}
}

\newcommand{\promptlabel}[1]{\noindent\textbf{\color{blue!50!black}#1}}

\begin{document}

\title{Assertion-Aware Test Code Summarization with Large Language Models}

\author{
\IEEEauthorblockN{Anamul Haque Mollah}
\IEEEauthorblockA{
\textit{University of North Texas}\\
Denton, TX, USA\\
AnamulMollah@my.unt.edu
}
\and
\IEEEauthorblockN{Ahmed Aljohani}
\IEEEauthorblockA{
\textit{University of North Texas}\\
Denton, TX, USA\\
AhmedAljohani@my.unt.edu
}
\and
\IEEEauthorblockN{Hyunsook Do}
\IEEEauthorblockA{
\textit{University of North Texas}\\
Denton, TX, USA\\
Hyunsook.Do@unt.edu
}
}

\maketitle

\begin{abstract}
Unit tests often lack concise summaries that convey test intent, especially in auto-generated or poorly documented codebases. Large Language Models (LLMs) offer a promising solution, but their effectiveness depends heavily on how they are prompted. Unlike generic code summarization, test-code summarization poses distinct challenges because test methods validate expected behavior through assertions rather than implementing functionality. This paper presents a new benchmark of 91 real-world Java test cases paired with developer-written summaries and conducts a controlled ablation study to investigate how test code-related components-such as the method under test (MUT), assertion messages, and assertion semantics-affect the performance of LLM-generated test summaries. We evaluate four code LLMs (Codex, Codestral, DeepSeek, and Qwen-Coder) across seven prompt configurations using n-gram metrics (BLEU, ROUGE-L, METEOR), semantic similarity (BERTScore), and LLM-based evaluation. Results show that prompting with assertion semantics improves summary quality by an average of 0.10 points (2.3\%) over full MUT context (4.45 vs. 4.35) while requiring fewer input tokens. Codex and Qwen-Coder achieve the highest alignment with human-written summaries, while DeepSeek underperforms despite high lexical overlap. The replication package is publicly available at \url{https://doi.org/10.5281/zenodo.17067550}
\end{abstract}

\begin{IEEEkeywords}
Unit Testing, Test Code Summarization, Assertion Messages, Large Language Models (LLMs)
\end{IEEEkeywords}

\section{Introduction}
\label{sec:Introduction}
Unit tests are a critical part of the codebase; they help ensure the correctness of the Method Under Test (MUT), prevent regressions, and document expected behavior. Yet, as test suites grow in size and complexity, understanding individual test cases—especially without comments—can become time-consuming. Developers often rely on summaries or comments to quickly understand a test’s intent, but such documentation is inconsistently written or entirely missing in many codebases.

Prior work~\cite{aljedaani2021test} has shown that developers often mitigate issues like vague tests or obscure behavior by writing descriptive test names or \textit{clear test comments}~\cite{spadini2020investigating}. These high-level summaries help developers understand the rationale behind a test without reading the entire test code. Despite their usefulness, test comments are often missing, outdated, or inconsistent—especially in auto-generated test suites from search-based approaches like EvoSuite~\cite{fraser2011evosuite} and transformer-based models~\cite{tufano2020unit, aljohani2024fine}.

To address this, several tools have been proposed to automatically generate summaries for test cases. TestDescriber, introduced by Panichella et al.~\cite{panichella2016impact}, uses a template-based strategy grounded in the SWUM (Software Word Usage Model) representation of code statements. Similarly, DeepTC-Enhancer, proposed by Roy et al.~\cite{roy2020deeptc}, combines deep learning with fixed templates to produce step-by-step natural language descriptions of tests. While both approaches improve test readability to some extent, they often produce verbose and redundant outputs that fall short of developer expectations for concise, high-quality summaries.


Large language models (LLMs) have recently emerged as promising tools for code comprehension~\cite{sun2024source}. When carefully prompted, models such as GPT-4 can produce fluent, context-appropriate text. Unlike production methods, test methods encode expected behavior through assertion constructs and often reference external methods under test. Summarizing them therefore requires reasoning over validation intent rather than implementation logic, making test-code summarization a distinct challenge compared to general code summarization. Nevertheless, most existing summarization benchmarks, such as CodeXGLUE~\cite{lu2021codexglue}, and many LLM-based techniques~\cite{jiang2024survey}, focus primarily on production code, overlooking the linguistic and structural nuances of test methods. Developer-written test summaries often begin with behavioral cues like “tests whether” or “verifies that,” and test bodies encode intent through assertions that specify expected outcomes.

In the context of test code, UTGen by Deljouyi et al.~\cite{deljouyi2024leveraging} demonstrated that large language models (LLMs) can enhance the readability of automatically generated unit tests, particularly through improvements in variable naming, test data selection, and embedded comments. However, their work does not provide empirical evidence regarding whether the generated test code summaries are meaningful or aligned with the original developer intent. Following this, Djajadi et al.~\cite{djajadi2025using} explored the capability of LLMs to generate high-quality summaries for test methods. Their study compared several prompting strategies, including baseline prompting, the inclusion of MUT context, and few-shot examples. The results indicated that prompt design plays a critical role in determining summary quality, influencing key dimensions such as content relevance, conciseness, and linguistic naturalness.
Despite recent advances, two key gaps remain. First, prior evaluations have been narrow in scale. For instance, Djajadi et al.~\cite{djajadi2025using} assessed LLM-generated summaries using only four test cases, which constrains the generalizability and robustness of their findings. Second, existing studies have largely neglected structural characteristics unique to test code—notably, assertion statements and their accompanying messages. These elements encode the intended behavior of the code and offer high-value semantic features for summarization. When such features are underemphasized, LLMs may fail to capture the behavioral intent of tests, particularly in long or complex cases~\cite{bavarian2022efficient}. Recent work~\cite{aljohani2025assertion, ahmed2024automatic} demonstrates that enriching prompts with semantic context can substantially enhance LLM performance in code summarization. While incorporating the full MUT context can provide additional behavioral information, it also increases token overhead and inference latency. In contrast, focusing on test code features offers a lightweight, scalable, and behaviorally rich alternative for improving summarization.
\noindent
To address these limitations, we propose the following research questions:
\begin{enumerate}
    \item \textbf{RQ1:} To what extent do assertion-level features influence the quality of LLM-generated summaries?
    \item \textbf{RQ2:} How do different code LLMs vary in their effectiveness across prompt variants?
\end{enumerate}

\noindent
This paper introduces a benchmark and ablation study for test code summarization. The benchmark comprises 91 developer-written Java tests, each paired with a gold comment written by the original developer. We evaluate seven prompt configurations that vary the inclusion of test code, MUT, assertion statements, and assertion semantics. We measure quality using n-gram metrics (BLEU, ROUGE-L, METEOR), semantic measures (BERTScore), and LLM-as-a-judge (LLM-Eval).

\noindent
This paper makes the following contributions:
\begin{enumerate}
  \item We introduce a benchmark of 91 developer-written Java test cases paired with developer-written comments.
  \item We conduct an ablation study across seven prompt configurations.
  \item We perform evaluation across four state-of-the-art code LLMs (Codex, Codestral, DeepSeek, and Qwen-Coder) and five evaluation metrics (BLEU, METEOR, ROUGE-L, BERTScore, and LLM Eval).
  \item We release a replication package to support reuse and future benchmarking.\footnote{\url{https://doi.org/10.5281/zenodo.17067550}}
\end{enumerate}

Our results provide actionable guidance on how to prompt code LLMs for test summarization. We show when MUT context helps, when assertion semantics can substitute for or complement MUT to reduce input length and cost, and how these components interact to improve content fidelity and conciseness in the generated summaries.

\noindent
The remainder of this paper is organized as follows: Section~\ref{sec:RelatedWork} reviews related work on automated test code summarization. Section~\ref{sec:ResearchMethod} outlines the methodology and experimental design. Section~\ref{sec:StudyResults} presents the results and key findings derived from our evaluation. Section~\ref{sec:Study_Implication} discusses the broader implications of the findings and provides additional insights. Section~\ref{sec:ThreatsToValidity} outlines the potential limitations and threats to the validity of our study. Finally, Section~\ref{sec:Conclusion} concludes the paper and highlights directions for future work.

\section{Related Work} 
\label{sec:RelatedWork}

This section reviews studies on automated test code summarization, focusing on the role of Template-based, Deep learning, and Large Language Models (LLMs). 

Prior work has shown that comments enable developers to fix bugs faster, understand test intent, and navigate large codebases more efficiently. For example, Panichella et al.~\cite{panichella2016impact} proposed \textit{TestDescriber} that uses templates informed by SWUM representations to generate test case summaries, improving bug-fixing performance in controlled experiments. Similarly, Roy et al.~\cite{roy2020deeptc} introduce \textit{DeepTC-Enhancer}, which combines deep learning with rule-based templates to create high-level step-by-step explanations of test behavior. While both tools increase readability, they often fall short of developer expectations by producing verbose and redundant outputs.

Large Language Models (LLMs) have recently been applied to test generation and comprehension. For instance, Deljouyi et al.~\cite{deljouyi2024leveraging} enhances automatically generated test cases by refining test names, input data, variable names, and inline comments using LLMs. Although their approach (i.e., UTGen) improves the presentation of test cases, it does not assess whether LLM-generated outputs are informative or accurate. That is, UTGen assumes the usefulness of LLM output without evaluating its summarization quality or its alignment with developer intent.

Building on this work, Djajadi et al.~\cite{djajadi2025using} explored whether LLMs could generate concise, understandable summaries for unit tests. Their study compared several prompting techniques, including baseline prompting, method-under-test (MUT) inclusion, and few-shot learning, using both open-source and commercial LLMs such as CodeLlama and GPT-4o. Through a small-scale user study with 11 participants and four test cases, they found that prompt engineering produced richer content while few-shot prompting led to more concise outputs. However, their study was limited in scale and did not deeply examine domain-specific features like assertion statements and messages, which often capture the core behavioral intent of a test.


Our study addresses these gaps by introducing a curated benchmark of 91 real-world Java test cases, each paired with developer-written summaries. We conduct a controlled ablation study using seven prompt variants to evaluate the impact of different input contexts-specifically, the test code, MUT, and assertion semantics. Unlike prior work in general code summarization~\cite{djajadi2025using, bavarian2022efficient}, We hypothesize that assertion-level information offers a lightweight yet behaviorally rich context that can match or outperform full MUT context. 


\section{Research Method}
\label{sec:ResearchMethod}

\begin{figure*}[h!]
    \centering
\includegraphics[width=\textwidth]{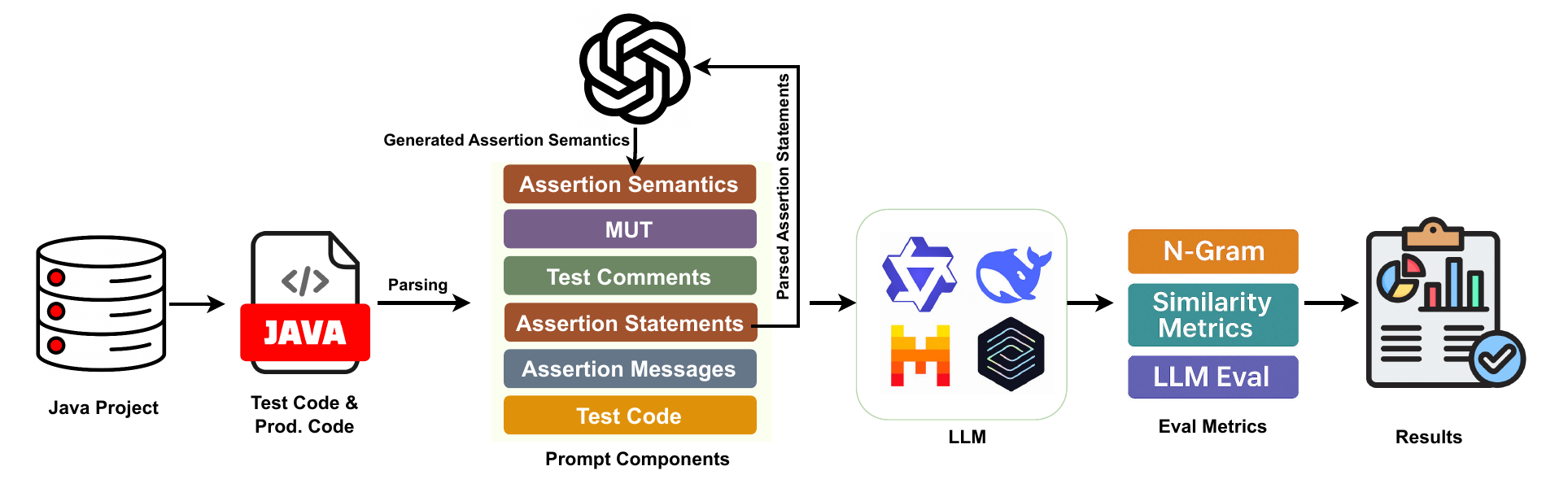}
\caption{Overall methodology for test code summarization. We parse Java test methods and extract their developer-written comments, assertion statements, and corresponding MUT. Each assertion is processed by an LLM to generate its semantic meaning. These parsed and enriched components then serve to construct structured prompts for our ablation study. Finally, we evaluate the quality of the generated summaries using standard text similarity and LLM-based metrics.}
    \label{fig:ApporachOverview}
\end{figure*}

Figure~\ref{fig:ApporachOverview} illustrates our methodology. It begins with \textit{Data Preparation}, where we filter Java test and production classes. We parse each test class to get the test code, test comments, assertion statements, assertion messages, and MUT. For assertion semantics, we used the parsed assertion statements for each test method and prompted GPT4o to generate the assertion semantics as described in Section~\ref{sec:ResearchMethod}.B. These preparations build the ablation study where we then infer the selected LLM to generate test summaries per ablation design. Finally, we used N-gram, similarity, semantic similarity, and LLM evaluation to measure the performance of the generated comments.

\vspace*{-3pt}
\subsection{Data Source}
\textbf{Benchmark:}
We built our dataset using CodeXGLUE~\cite{lu2021codexglue}, a large-scale benchmark suite for code-related tasks including code summarization, code search, clone detection, and defect prediction. It supports six programming languages: Java, Python, JavaScript, PHP, Ruby, and Go. Our focus is on the \textit{code summarization} task, which involves generating natural language descriptions from code snippets. However, CodeXGLUE does not provide a dedicated subset for test methods. To address this, we curated a new dataset by extracting test methods from Java projects included in the original CodeXGLUE corpus. We leveraged the open-source tools \texttt{TestFileDetector} and \texttt{TestFileMapping}~\cite{testfilemapping} to identify Java files containing unit tests and map each test file to its corresponding production class, enabling accurate identification of the Method Under Test (MUT).

Following the methodology by Takebayashi et al.~\cite{takebayashi2023exploratory}, we employed the JetBrains Software Development Kit (SDK)\footnote{\url{https://github.com/JetBrains/intellij-community}} to extract test classes. We selected classes that included at least one \texttt{@Test} annotation, ensuring that the extracted methods were valid test cases. We further filtered to retain only those test methods that contained assertions accompanied by descriptive assertion messages. For each selected method, we extracted the assertion statement, assertion message, MUT, and any associated code comments. This resulted in a total of 210 test methods.

\noindent
\textbf{Filtering and Preprocessing}
To ensure inclusion of only high-quality and meaningful test comments, we applied a series of filtering steps. Our goal was to preserve comments that explicitly convey the rationale behind test failures, while discarding vague, incomplete, or noisy comments—such as placeholder or non-informative comments (e.g., \texttt{TODO})~\cite{potdar2014exploratory}. Specifically, we excluded:

\begin{itemize}
\item Comments that were empty or written in non-English languages.
\item Comments with fewer than four words.
\item Placeholder comments containing only keywords like \texttt{TODO}, \texttt{FIXME}, or \texttt{deprecated}.
\item Comments that exclusively reference external resources (e.g., URLs, HTML tags, or GitHub links).
\end{itemize}

\noindent
After preprocessing, 91 out of the original 210 test methods remained and were used in our final evaluation.

\subsection{Ablation Study Design}
To understand how different parts of a prompt affect the quality of test summaries, we designed an ablation study. Our goal is to examine the individual and combined effects of four main components: the test method, assertion message, assertion semantics, and the Method Under Test (MUT). To generate assertion semantics, each extracted assertion was provided to GPT-4o, which produced a concise natural-language meaning. By adding or removing each part, we evaluate its contribution to summary generation. We created seven different prompt versions:

\begin{itemize}
    \item \textbf{Test method only}: The baseline prompt, containing only the raw Java test method extracted from the codebase.
    
    \item \textbf{Test method with assertion messages}: Includes the test method along with assertion messages for all assertion statements.

    \item \textbf{Test method without assertion messages}: Consists of test method where none of the assertion statements contain any messages.

    \item \textbf{Test method with assertion semantics}: Enhances the test method by providing each test with its natural language interpretation, while omitting both assertion messages and MUT content.

    \item \textbf{Test method with assertion messages and semantics}: Includes assertion messages and semantic meanings for all assertion semantics.

    \item \textbf{Test method with MUT}: Combines the test method with MUT, without any assertion message or semantic descriptions.
    
    \item \textbf{Test method with assertion messages, semantics and MUT}: Enriches the test method with three layers of augmentation: assertion messages, MUT, and natural language semantic explanations for each assertion.

\end{itemize}

\noindent 
Most components, such as the test method, assertion message, and MUT, were extracted directly from the test or its associated production code. However, the assertion semantics were not present in the code itself and had to be generated separately. To obtain these semantics, we first isolated each assertion statement from its corresponding test method. We then applied an instruction-based prompting technique, following the approach of \cite{deljouyi2024leveraging, assertify2024}, using GPT-4o to transform the assertion into a concise natural-language description that captures its intent. This semantic explanation conveys what the test is verifying or why the assertion might fail.
For instance, the assertion
\texttt{assertThat("Invalid age", user.getAge(), is(equalTo(18)));}
was rewritten semantically as:
“Checks that the user’s age equals 18.”

\subsection{Prompt Design}
We employed a unified, role-based instruction prompt template across all selected large language models (e.g., DeepSeek, Qwen-Coder, Codestral, Codex-Mini). Prior studies have demonstrated that role-based prompting and task-specific instructions can enhance model alignment and output quality~\cite{zheng2023helpful}. In designing our prompts, we incorporated techniques from recent work~\cite{deljouyi2024leveraging, sun2024source}, which highlight the effectiveness of structured prompting in improving LLM performance on code summarization.
Our final prompt template is described as follows:


\vspace*{2pt}
\begin{quote}\small\ttfamily
\textbf{Role}: You are a senior Java test engineer focused on summarizing unit-test behavior and expected outcomes.

\textbf{Instruction}:  
Your task is to analyze the provided test code and generate a short summary comment.

\vspace{0.5\baselineskip}

[CODE]  
...test method ...  
[/CODE]

\vspace{0.5\baselineskip}
Please find more information about assertions in \texttt{[ASSERTIONS][/ASSERTIONS]} and methods under test in \texttt{[MUTS][/MUTS]}

\vspace{0.5\baselineskip}

[ASSERTIONS]  
...assertions context...
[/ASSERTIONS]

\vspace{0.5\baselineskip}

[MUTS]  
...MUT...  
[/MUTS]\\
Please generate a short summary comment in one sentence for the test code.  
Please do not use more than 20 words.

\vspace{0.5\baselineskip}
[SUMMARY]  
...comment...  
[/SUMMARY]
\end{quote}

\noindent
In our prompt, the test method was always placed inside the \texttt{[CODE]...[/CODE]} block, serving as the primary input to the model. For the remaining ablation study, we varied the prompt using the features described in Section~\ref{sec:ResearchMethod}.B.

\noindent
In our dataset, developer-written comments average 14 words, with a maximum of 40; notably, 43\% exceed the average, reflecting substantial variation in length. This variability supports the use of a fixed upper bound to promote consistency and conciseness in generated summaries. To enforce this constraint, we instructed the model with the explicit guideline: \textit{“Please do not use more than 20 words,”} aligning with recent findings that length limits improve summary quality~\cite{sridhara2024icing}. We also adopted a concise, behavior-focused instruction~\cite{sun2024source} as follows: \textit{“Please generate a short summary comment in one sentence for the test code.”}

\subsection{Code Models and Inference}
We evaluate four instruction-tuned code generation models commonly used in recent software engineering research:

\begin{itemize}
\item Qwen2.5-Coder-32B-Instruct \cite{hui2024qwen2}\footnote{\url{https://huggingface.co/Qwen/Qwen2.5-Coder-32B-Instruct}}
\item Codestral-22B-v0.1 \cite{mistral2024codestral}\footnote{\url{https://huggingface.co/mistralai/Codestral-22B-v0.1}}
\item DeepSeek-Coder-33B-Instruct \cite{guo2024deepseek}\footnote{\url{https://huggingface.co/deepseek-ai/deepseek-coder-33b-instruct}}
\item Codex (\texttt{Codex-Mini})\footnote{\url{https://github.com/openai/codex}}
\end{itemize}

\noindent 
Model selection was based on their widespread adoption in recent LLM for code studies\cite{zhao2023survey}, as well as strong performance on code-related benchmarks such as EvalPlus~\cite{liu2023your}. For inference, we used each model’s default configuration on Hugging Face or API, preserving the model’s intended decoding behavior.

\subsection{Evaluation}
We use BLEU~\cite{papineni2002bleu}, METEOR~\cite{banerjee2005meteor}, and ROUGE-L~\cite{lin2004rouge} to measure n-gram overlap between generated and reference comments. To address the limitations of purely lexical metrics~\cite{haque2022semantic}, we also employ BERTScore~\cite{zhang2019bertscore} to capture deeper semantic similarity. Moreover, recent studies~\cite{sun2023automatic, su2024distilled, wu2025can} have shown that LLM-generated text can often surpass human-written references in quality. This raises concerns about the suitability of relying solely on reference-based similarity metrics. To address this, we adopt an LLM-based evaluation to rank the quality of generated comments. This method captures subjective elements such as usefulness and clarity, which are not easily measured via automatic metrics. Following the approach by Sun et al.~\cite{sun2024source}, we designed a structured prompt to guide the evaluation:

\begin{promptboxx}
  \promptlabel{ LLM Evaluation Prompt}\\

  Here is a piece of Java unit-test code and ONE comment. Please rate the comment on a scale from 1 to 5, where a higher score indicates better quality.\\

  Consider the following: \\
  1) Accurately states what behavior is verified and under which conditions.\\
  2) Reflects the expected outcomes expressed by the assertions (values, state changes, exceptions).\\
  3) Identifies the method, class, feature, or scenario under test.\\
  4) Mentions edge cases, negative paths, exceptions, or side effects when applicable (no penalty if not applicable).\\
  5) Is expressed naturally and concisely, without burdening the developer with reading.\\
  6) Helps the developer understand the code quickly.\\

  Return ONE line ONLY in the exact format: \\
  \texttt{Score: X} \\
  (where X is a number from 1 to 5, integer or decimal).\\

  \vspace{0.3\baselineskip}
  \codeblock{{Code:} \{code\_block\}}
  \vspace{0.3\baselineskip}
  \codeblock{{Comment:} \{comment\_block\}}
  \vspace{0.3\baselineskip}
\end{promptboxx}

\vspace*{5pt}
The original evaluation prompt proposed by~\cite{sun2024source} was designed to assess the usefulness of generated source code comments. In our study, we modified this prompt to better 
suit the context of test code comments.

\vspace{0.5em}
\begin{figure}[H]
\centering
\begin{minipage}{\linewidth}
\begin{lstlisting}[language=Java, basicstyle=\ttfamily\small, caption={Java test method with developer written comment}, label={lst:SummaryComparison}]
/* Test that the principal ID assigned to a principal is correctly stored and 
   returned using the Java Principal API */
void principalIdTest() {
    final String USER_ID = "someId";
    Principal principal = new UserPrincipal(USER_ID);
    String error = "Principal ID does not match expected value";
    assertThat(error, principal.getName(), is(equalTo(USER_ID)));
}
\end{lstlisting}
\end{minipage}
\vspace{-0.5em}
\end{figure}

\noindent The following examples present Codex and Codestral generated summaries under different prompt variants shown in Listing~\ref{lst:SummaryComparison}.\\

\noindent
\textbf{Codex summaries:}\par
\noindent\textbf{Test method only:} Verifies that the Principal’s ID matches the expected user ID using an assertion. \par
\noindent\textbf{Test method with assertion messages:} Verifies that the Principal’s ID matches the expected user ID ``someId.'' \par
\noindent\textbf{Test method without assertion messages:} Verifies that UserPrincipal’s getName() returns the correct user ID. \par
\noindent\textbf{Test method with assertion semantics:} Verifies that the Principal’s name matches the expected user ID ``someId.'' \par
\noindent\textbf{Test method with assertion messages and semantics:} Verifies that the Principal’s name matches the expected user ID ``someId.'' \par
\noindent\textbf{Test method with MUT:} Verifies UserPrincipal's getName() returns the correct ID, ensuring the constructor correctly assigns the name. \par
\noindent\textbf{Test method with assertion messages, semantics, and MUT:} Tests if UserPrincipal correctly returns the user ID as its name. \par

\vspace{1em}
\noindent
\textbf{Codestral summaries:}\par
\noindent\textbf{Test method only:} The test verifies that the Principal’s ID matches the expected user ID. \par
\noindent\textbf{Test method with assertion messages:} The test ensures the correctness of the Principal's ID. \par
\noindent\textbf{Test method without assertion messages:} The test verifies that the UserPrincipal's name matches the provided USER ID. \par
\noindent\textbf{Test method with assertion semantics:} The test code verifies that the principal's name matches the expected user ID. \par
\noindent\textbf{Test method with assertion messages and semantics:} The test verifies that the principal's name matches the expected user ID. \par
\noindent\textbf{Test method with MUT:}The test verifies that the UserPrincipal's getName() method correctly returns the provided user ID.\par
\noindent\textbf{Test method with assertion messages, semantics, and MUT:} The test verifies that the 'getName' method of 'UserPrincipal' correctly returns the provided user ID. \par

\section{Results}
\label{sec:StudyResults}
This section presents the performance of various large language models (LLMs) across different prompt configurations for test code summarization. We report results using both lexical metrics (e.g., BLEU, ROUGE, METEOR) and semantic evaluation (LLM Eval) to assess the quality and alignment of generated summaries with human-written ground truths. Our analysis emphasizes the effect of the test code components such as including assertion messages, semantics, or MUT on summary quality. 
In Listing~\ref{lst:SummaryComparison}, we illustrate concrete examples of test comments generated by Codex and Codestral against the corresponding developer-written ground truth comment.

\begin{table*}[t]
\caption{
\textbf{Performance of LLMs in Generating Test Summaries across Different Variants.}
The average ground-truth LLM-Eval score is 3.43/5. All n-gram overlap metric scores are measured using the method described in~\cite{zhang2020retrieval, sun2024source}. Bolded values indicate the highest per-model score for each metric.
}
\centering
\resizebox{\linewidth}{!}{
\begin{tabular}{llccccc}
\hline
\textbf{Model} & \textbf{Variant} & \textbf{BLEU} & \textbf{METEOR} & \textbf{ROUGE-L} & \textbf{BERTScore F1} & \textbf{LLM Evals (out of 5)} \\
\hline
\multirow{7}{*}{Codex} 
 & Test method only & 17.46 & 27.53 & 21.90 & 86.62 & 4.74 \\
 & Test method with assert msg & 17.24 & 25.88 & 22.26 & 86.76 & 4.82 \\
 & Test method without assert msg & 17.56 & 27.95 & 21.46 & 86.55 & 4.76 \\
 & Test method with semantics & \textbf{17.77} & 28.36 & \textbf{22.56} & \textbf{86.76} & 4.81 \\
 & Test method with assert msg, semantics & 17.33 & 27.25 & 22.38 & 86.67 & \textbf{4.85} \\
 & Test method with mut & 17.55 & \textbf{29.04} & 21.30 & 86.50 & 4.78 \\
 & Test method with assert msg, mut, semantics & 16.95 & 27.64 & 21.95 & 86.64 & 4.81 \\
\hline
\multirow{7}{*}{Codestral} 
 & Test method only & 18.64 & 31.82 & \textbf{26.19} & 86.91 & 4.48 \\
 & Test method with assert msg & 17.91 & 28.41 & 25.46 & 86.97 & 4.64 \\
 & Test method without assert msg & \textbf{19.24} & 31.97 & 25.87 & \textbf{87.03} & 4.57 \\
 & Test method with semantics & 17.95 & 31.53 & 24.52 & 86.72 & 4.68 \\
 & Test method with assert msg, semantics & 18.11 & 31.45 & 25.09 & 86.80 & \textbf{4.83} \\
 & Test method with mut & 17.86 & \textbf{33.90} & 25.38 & 86.67 & 4.58 \\
 & Test method with assert msg, mut, semantics & 17.68 & 31.40 & 24.76 & 86.72 & \textbf{4.83} \\
\hline
\multirow{7}{*}{DeepSeek} 
 & Test method only & 15.81 & 22.21 & 22.87 & 87.37 & 3.08 \\
 & Test method with assert msg & 15.71 & 21.75 & 23.33 & \textbf{87.42} & 3.17 \\
 & Test method without assert msg & 15.74 & 20.84 & 22.75 & 87.14 & 3.10 \\
 & Test method with semantics & 16.48 & 23.89 & 23.31 & 87.37 & 3.54 \\
 & Test method with assert msg, semantics & \textbf{17.45} & 25.41 & \textbf{24.17} & \textbf{87.46} & 3.74 \\
 & Test method with mut & 17.27 & \textbf{25.45} & 22.83 & 87.41 & 3.38 \\
 & Test method with assert msg, mut, semantics & 17.25 & 25.16 & 23.00 & 87.15 & \textbf{3.88} \\
\hline
\multirow{7}{*}{Qwen-Coder} 
 & Test method only & 17.91 & 28.64 & 23.41 & 86.92 & 4.60 \\
 & Test method with assert msg & 17.28 & 26.36 & 24.21 & 87.01 & 4.70 \\
 & Test method without assert msg & \textbf{18.21} & \textbf{28.71} & \textbf{24.80} & \textbf{87.16} & 4.63 \\
 & Test method with semantics & 16.57 & 25.02 & 21.80 & 86.65 & 4.75 \\
 & Test method with assert msg, semantics & 17.38 & 26.03 & 23.14 & 86.80 & 4.75 \\
 & Test method with mut & 17.95 & 28.24 & 23.78 & 86.86 & 4.66 \\
 & Test method with assert msg, mut, semantics & 17.50 & 26.80 & 23.42 & 86.78 & \textbf{4.76} \\
\hline
\end{tabular}
}
\label{table:performance-models}
\end{table*}

\subsection{RQ1: Effect of Assertion-Level Features on Summary Quality}
We examine the effectiveness of different prompt components, specifically, \textit{assertion messages}, \textit{assertion semantics}, and \textit{method-under-test (MUT)} on the quality of LLM-generated test summaries. Table~\ref{table:performance-models} reports results for all prompt variants across lexical metrics (BLEU, ROUGE-L, METEOR), the semantic metric (BERTScore), and human-aligned LLM evaluation scores. While developer-written summaries serve as ground truth for lexical and semantic metrics, LLM-Eval is an independent quality assessment in which both developer and model summaries were rated by GPT-4o using the same rubric. As a result, models can occasionally score higher than the developer reference if their summaries are judged clearer or more concise. Under this metric, most models surpassed the developer-written baseline (LLM-Eval = 3.43/5), except for DeepSeekCoder, whose outputs consistently fell short across all variants.

Assertion-enhanced prompts, especially those that include both assertion messages and their semantic interpretations—consistently led to the best performance across models. For example, Codex reached its highest LLM Eval score (4.85) and showed strong results on lexical metrics (BLEU = 17.77, ROUGE-L = 22.56, BERTScore = 86.76) when both assertion components were included. DeepSeek also improved noticeably with this setup: its BLEU score rose from 15.81 (with test code only) to 17.45, and ROUGE-L increased from 22.87 to 24.17. Similar trends were seen in Codestral and Qwen-Coder, both achieving their top LLM Eval scores (4.83 and 4.76, respectively) when assertion-related information was added. Interestingly, Qwen-Coder performed best on BLEU and METEOR when using the variant without explicit assertion messages—suggesting that the model might still benefit from structural patterns or implicit cues even in the absence of direct assertion content.
In contrast, MUT-only variants produced mixed results. While they occasionally yielded strong lexical overlap—for instance, METEOR scores for Codex (29.04) and Codestral (33.90)-their LLM Eval scores plateaued or declined (Codex = 4.78, Codestral = 4.58). This suggests that although MUT context can enhance surface-level token alignment, it often lacks the focused behavioral semantics needed for high-quality summarization. Similarly, in DeepSeek, MUT-only prompts improved over the baseline (LLM Eval = 3.38 vs. 3.08), but remained lower than assertion-rich prompts like the configuration combining assertion messages, MUTs, and semantics (3.88). For Qwen-Coder, MUT yielded a decent 4.66, but still trailed assertion-enhanced prompts. These findings suggest that while the MUT provides some behavioral signal, it tends to introduce verbosity without consistent improvements in semantic alignment. Moreover, prompts that include all available test context—that is, the test code, MUT, assertion messages, and semantic summaries—did not always outperform simpler configurations. For Qwen-Coder, this full-context setup scored 4.76, nearly identical to the variant using only assertion messages and semantics (4.75), indicating diminishing returns when layering all components. Overall, while the MUT adds some value, assertion-level features are more targeted and consistently more effective in guiding LLMs toward concise, behaviorally accurate summaries.

When it comes to Test-only prompts, it consistently underperformed across models. Codex (4.74), Qwen‑Coder (4.60), and Codestral (4.48) all showed weaker human‑aligned scores when prompts excluded assertion cues or MUT context. The most notable drop occurred in DeepSeek, where the test‑only variant reached just 3.08—substantially below the overall human baseline. These results confirm that relying solely on the test method, without behavioral or structural signals, leads to summaries that are often vague and poorly aligned with developer expectations.

Finally, we observe that different models respond differently to prompt variants. Codex and Qwen-Coder maintained strong performance across both lexical and human-aligned metrics, showing robustness to prompt composition. Codestral, by contrast, excelled in BLEU and ROUGE-L but lagged in LLM Eval scores, indicating a disconnect between n-gram similarity and human preference. DeepSeek consistently underperformed in LLM Eval despite competitive BERTScores. 

\vspace*{5pt}
\begin{mdframed}[linewidth=1pt, linecolor=gray!75!black, backgroundcolor=gray!5, roundcorner=5pt]
\textbf{Summary of RQ1:}  
Assertion messages and semantics consistently produced the highest-quality summaries across all models. Test-only variants performed worst, while MUT-only prompts showed limited gains—improving lexical overlap but not human-aligned scores. Full-context prompts offered minimal improvement over concise, assertion-focused inputs.
\end{mdframed}

\vspace*{2pt}
\subsection{RQ2: Cross-Model Comparison of Summary Quality}
We now examine how different LLMs compare in their ability to generate high-quality test summaries across all prompt variants. Our analysis draws on both automatic metrics (BLEU, METEOR, ROUGE-L, BERTScore) and human-aligned LLM Eval scores. The goal is to identify which models consistently perform well, which struggle, and how they respond to prompt design.

Codex demonstrated the strongest alignment with human preferences, achieving the highest LLM Eval score of 4.85 when enhanced with assertion messages and semantics. It also performed well on lexical metrics (e.g., BLEU = 17.77, ROUGE-L = 22.56), indicating that it generated summaries that were both semantically and behaviorally aligned. Across prompt variants, Codex responded reliably to assertion-related cues, maintaining top-tier results even with concise prompts.

Qwen-Coder also exhibited robust and stable performance. It attained a competitive LLM Eval score of 4.76, alongside the highest BLEU (18.21) and ROUGE-L (24.80) scores among its configurations, particularly when assertion information was excluded—implying that Qwen-Coder may implicitly capture assertion intent from code structure alone. Its responsiveness to both detailed and minimal prompts shows adaptability, though its LLM-Eval scores remained slightly lower than Codex’s.

Codestral, in contrast, showed a clear mismatch between lexical similarity and human evaluation. It achieved the highest BLEU (19.24) and ROUGE-L (26.19) scores of all models, especially in test-only and test-without-assertion variants. However, its LLM Eval scores plateaued at 4.83, suggesting that its summaries, while textually similar, may be overly verbose or lacking the behavioral details. 

DeepSeekCoder consistently ranked lowest in human evaluations. Despite reaching a BERTScore of 87.46 and showing modest improvements in BLEU and METEOR with assertion-rich prompts, its highest LLM Eval score peaked at just 3.88, well below both the Codex/Qwen range and the ground truth comment average (3.43). These results suggest DeepSeek captures surface semantics but fails to distill behavioral intent into concise and meaningful summaries.

\vspace*{5pt}
\begin{mdframed}[linewidth=1pt, linecolor=gray!75!black, backgroundcolor=gray!5, roundcorner=5pt]
\textbf{Summary of RQ2:} 
Codex and Qwen-Coder emerge as the most reliable models for test code summarization, particularly when guided by behaviorally rich prompts. Codestral excels in token-level overlap but may lack human-preferred clarity. DeepSeekCoder struggles to meet human standards despite preserving semantic content.
\end{mdframed}

\section{Additional Discussions and Implications} 
\label{sec:Study_Implication}
Across models is the value of incorporating test behavioral context—especially through assertion semantics. Prompts that included these semantic cues consistently led to higher-quality summaries. Notably, Codex and Qwen-Coder performed nearly as well, or even better, with just assertion semantics than they did when provided with the full MUT. This suggests that concise, behavior-focused input helps models generate clearer and more useful summaries than verbose structural context. Semantic cues appear to help the model capture what is being tested rather than how, which better aligns with how developers interpret test cases.

Building on this, we found that assertion components—whether messages or semantics—are central to summary quality. Across all models, prompts enriched with assertion information outperformed test-only variants in both lexical and human-aligned metrics. The absence of assertion messages, even when other context was present, led to weaker summaries that often lacked intent or behavioral clarity. These results reinforce the role of assertions as the core signal in unit tests. Instruction-tuned LLMs benefit from focused prompts that surface these behavioral cues. 

However, including too many components in the prompt can backfire. While it might seem intuitive to supply the model with all available information: test code, assertions, semantics, and the MUT—our results show that this “full-context” approach doesn’t always help. In some cases, such as with Qwen-Coder and Codestral, these overloaded prompts offered no improvements or even slightly worse performance than more concise assertion-based prompts. This suggests that excessive or loosely related information can distract the model, leading to diminished returns. 

Finally, our analysis highlights a key concern in evaluation methodology. We observed several cases where traditional n-gram-based metrics like BLEU and ROUGE gave high scores to summaries that, upon human inspection, were vague or misaligned. For instance, Codestral’s test-only variant received strong lexical scores but was rated poorly in LLM Eval. This discrepancy shows that word overlap alone cannot capture summary quality. As LLMs become more central to code summarization, future evaluations must place greater emphasis on semantic fidelity and human-aligned assessments to avoid misleading conclusions.

\section{Threats to Validity}
\label{sec:ThreatsToValidity}

\textbf{External Validity:}
The generalizability of our findings is limited by the scope of our dataset and experimental design. Specifically, we focus exclusively on Java unit tests written using the JUnit4 framework. As a result, our conclusions may not directly apply to other testing frameworks such as JUnit5 or TestNG, which offer different structural features (e.g., dynamic tests, parameterized tests, and distinct lifecycle annotations) that may influence summarization effectiveness. Additionally, our main dataset consists of test methods that contain assertion messages, which may introduce a bias toward well-documented or better-structured tests. However, to address this concern, we included test methods without assertion messages in our ablation study, allowing us to evaluate the model’s performance across a broader spectrum of real-world test code.

\vspace*{3pt}
\textbf{Internal Validity:}
Our study relies on GPT‑4o in two capacities: (1) to generate natural language semantics for assertion statements, and (2) to provide human‑aligned quality judgments through LLM‑Eval. These dual roles introduce risks of model bias—both in the content being evaluated and in the evaluation itself—due to potential inconsistencies or subjective tendencies in generative models. To reduce these threats, we adopted the LLM‑Eval configuration from Sun et al.~\cite{sun2024source} and applied identical decoding settings. Additionally, we manually reviewed a sample of generated semantics and evaluations to ensure correctness and coherence, and we grounded our methodology in recent validated practices where LLMs have been successfully used for similar tasks~\cite{tafreshipour2024prompting, pister2024promptset, sun2024source}. Furthermore, all model inferences were performed using the default hyperparameters provided by each model’s Hugging Face repository, without additional fine-tuning. While this ensured consistency and fairness across models, it may not reflect each model’s optimal performance. Lastly, we limited our study to instruction-tuned LLMs, excluding base models. Although this decision aligns with practical usage scenarios, future work may explore base or custom fine-tuned models specifically adapted for software engineering tasks such as test summarization or assertion message generation.

\section{Conclusions}
\label{sec:Conclusion}
This study presents a systematic evaluation of test code summarization using large language models. We introduce a benchmark of 91 developer-written Java test cases and perform an ablation study across seven prompt configurations and four LLMs. Our results reveal that structural prompt components-especially assertion semantics-play a critical role in guiding models to produce accurate and concise test summaries. 
We find that assertion-level context often outperforms or complements full method-under-test (MUT) context, offering a more efficient alternative for summarization tasks. Among the models evaluated, Codex and Qwen-Coder consistently generate the most natural and meaningful summaries, while DeepSeek underperforms despite achieving high token overlap on standard metrics. 
Our findings provide practical guidance for prompt design in test summarization, and highlight the value of lightweight behavioral components. We release our benchmark and evaluation pipeline to support further work in test code comprehension and LLM-based documentation tools.

\bibliographystyle{IEEEtran}
\bibliography{Bib/bib}

\end{document}